\documentclass{jaa}
\usepackage{txfonts}
\usepackage{graphicx}
\usepackage{color}

\usepackage[authoryear]{natbib}

\newcommand{\be}{\begin{equation}}
\newcommand{\e}{\end{equation}}

\newcommand{\bear}{\begin{eqnarray}}
\newcommand{\ear}{\end{eqnarray}}

\def\th{\theta}

\def\apj{ApJ}

\def\mnras{MNRAS}

\def\u{{\vec U}}
\def\th{\vec{\theta}}

\def\E{\hat{E}}

\def\xh1{x_{{\rm H~{\sc I}\,}}}

\def\nat{Nature}
\def\HII{H~{\sc ii}\,}

\begin{document}
\title[Probing sources during reionization and cosmic dawn with SKA]
{Probing individual sources 
during reionization and cosmic dawn using SKA HI 21-cm observations} 
\author[ Datta et al] {\small{Kanan K. Datta$^{1}$
\thanks{e-mail: kanan.physics@presiuniv.ac.in}, Raghunath Ghara$^{2}$, Suman Majumdar$^{3}$, T. Roy Choudhury$^{2}$,}\\
 {Somnath Bharadwaj$^4$, Himadri Roy$^{1}$, Abhirup Datta$^5$ }
\\
 \tiny{$^1$Department of Physics, Presidency University, Kolkata, W.B, India}\\
 \tiny{$^2$National Centre for Radio Astrophysics, TIFR, Post Bag 3, Ganeshkhind, Pune 411007, India}\\
 \tiny{$^3$Department of Physics, Blackett Laboratory, Imperial College, London SW7 2AZ, UK}\\
 \tiny{$^4$Department of Physics \& Centre for Theoretical Studies, Indian Institute of Technology Kharagpur, Kharagpur - 721302, India}\\
 \tiny{$^5$Centre for Astronomy, Indian Institute of Technology Indore, Indore - 452020, India}
}
\date {}
\maketitle
\label{firstpage}
\begin{abstract} 
\small{Detection of individual luminous sources during the reionization epoch and cosmic dawn through 
their signatures in the HI 21-cm signal is one of the direct approaches to probe the epoch.
Here, we summarize our previous works on this  and present preliminary results on the prospects of 
detecting such sources using the SKA1-low experiment. We first discuss the expected HI 21-cm
signal around luminous sources at different stages of reionization and cosmic dawn. 
We then introduce  two visibility based estimators for detecting such signal: one based on the matched filtering
technique and the other relies on simply combing the visibility signal from different baselines and frequency channels.

We find that that the SKA1-low should be able to detect ionized bubbles of radius $R_b \gtrsim 10$ Mpc 
with $\sim 100$ hr of observations at redshift $z \sim 8$ provided that the mean outside 
neutral Hydrogen fraction $\xh1 \gtrsim 0.5$. 
We also investigate the possibility of detecting HII regions around known bright QSOs such as around 
ULASJ1120+0641 discovered by \citet{mortlock11}.  We find that a $5 \sigma$ detection 
is possible with $600$ hr of SKA1-low observations if the QSO age and the outside $\xh1$ 
are at least $\sim 2 \times 10^7$ Myr and $\sim 0.2$ respectively.
 
Finally, we investigate the possibility of detecting the very first X-ray and Ly-$\alpha$ sources
during the cosmic dawn. We consider mini-QSOs like sources which emits in X-ray frequency band. We find 
that with  a total $\sim 1000$ hr of observations, SKA1-low  should be able to detect those sources 
individually with a $\sim 9 \sigma$ significance at redshift $z=15$. 
We  summarize how the SNR changes with various parameters related to the 
source properties.}

\end{abstract}

\section{Introduction}
The emergence of first galaxies, quasars in the Universe is one of the significant
events in its history. In the standard scenario, the Ly-$\alpha$, X-ray, UV photons
produced by these first sources percolated through the intergalactic medium
(IGM) and completely altered its thermal and ionization state. Unfortunately, we know very 
little about this landmark event and nature and properties of first sources .

Recently, tens of extremely bright quasars at redshifts $z \gtrsim 6$ have been detected 
by various surveys \citep{fan06, mortlock11, venemans15}. In addition, hundreds of high 
redshift galaxies at similar redshifts have been discovered by various telescopes  
\citep{ouchi10, hu10, kashikawa11, ellis13, bouwens15}. These sources 
are expected to have played an important role during the reionization epoch. We 
expect many such sources to exist even at higher redshifts during the initial stage of 
reionization and cosmic dawn.   

It is likely that the UV photons from these sources ionized the neutral Hydrogen (H I) atoms 
around them and created H II regions (ionized bubbles) which are embedded in H I medium. 
The Ly-$\alpha$  from the very first luminous sources during the cosmic dawn coupled the 
IGM temperature with HI spin temperature. Similarly the X-ray photons from X-ray sources 
heated up the IGM and therefore raised both the IGM kinetic and the HI spin 
temperature. The coupling and the heating are initially efficient near the sources.  Previous studies showed 
that it is possible to detect the 21-cm signatures around these individual sources 
using current low-frequency telescopes \citep{datta07, geil08, datta3, datta09, malloy13} 
which would be useful in constraining the IGM and source properties \citep{majumdar12,datta12}. 
We also expect that upcoming space telescopes such as the James Webb Space Telescope (JWST)\footnote{http://jwst.nasa.gov} 
should also be able to detect some of these brightest sources during the cosmic dawn and 
reionization epoch in the optical/infrared band \citep{zackrisson11, desouza13, desouza14}. 

One of the straightforward approaches to understand the reionization epoch and cosmic dawn 
is to detect individual luminous sources through their signature on the HI 21-cm signal. 
Existing low frequency radio telescopes such as the GMRT, MWA, LOFAR primarily aim to detect the HI 
21-cm signal statistically by measuring quantities such as the power spectrum, rms, skewness of 
the HI 21-cm brightness temperature fluctuations. Here, we take an alternative but direct approach to 
explore the reionization epoch and cosmic dawn 
through detection of HI 21-cm signature around individual sources. In this paper, our aim is to 
summarise our previous works on this issue and present some preliminary results on the prospects of detecting 
such sources using the SKA1-low experiment  which is the low frequency part 
of the Square Kilometer Telescope \footnote{https://www.skatelescope.org/} to be built in phase 1.  
With much higher number antenna, better baseline coverage and detectors, the SKA1-low is expected to be 
much more sensitive telescope for detecting such objects individually with much less observation 
time \citep{ghara16, mellema12, mellema15}. After a brief discussion on nature of the expected 
HI 21-cm signal around luminous sources at different 
stages of reionization and cosmic dawn, we introduce two visibility based methods for 
detecting such signal:  (i) the first method is based on the matched filtering principle which was first introduced 
in \cite{datta07} and explored in detail in subsequent works (\cite{datta3, datta09, datta12}, \cite{majumdar11, majumdar12}, 
\cite{malloy13} ) and (ii) the second method relies on 
simply combing the visibility signal from different baselines and frequency channels (\cite{ghara15a}). 

The outline of the article is as follows: In section 2, we discuss the HI 21-cm signal profile 
around individual sources during the cosmic dawn and reionization epoch. In subsection 2.3, 
we calculate and discuss corresponding visibility signal. Section 3 discusses two visibility based 
estimators which have been developed for detecting the signal discussed in section 2. We also describe the 
filter we consider for the first estimator which uses the matched filter technique. 
Section 4 discusses the results on the detectability of ionized bubbles, known bright QSOs
in the reionization epoch and the first sources during the cosmic dawn using SKA1-low 
telescope. Section 5 presents a summary of the paper. 

\section{The HI 21-cm signal around individual sources}
The first sources are expected to emit in UV, Ly-$\alpha$, X-ray frequencies. In a likely 
scenario, a fraction of the Ly-$\alpha$ photons from the first generation of stars escape 
from their host environment and couple the H I spin temperature with the IGM gas kinetic
temperature very quickly. At the same time the soft X-ray photons emitting 
from sources like the mini-QSOs, X-ray binaries, Pop III 
stars etc. enter into the IGM and heat it up. This causes the HI spin 
temperature above the background CMB temperature. Subsequently, the UV photons 
start ionizing HI surrounding the sources and create ionized bubbles which are embedded 
into HI medium. The Ly-$\alpha$ coupling, X-ray heating, and UV ionization are initially
done near the sources and slowly spread over the entire IGM. We expect three
distinct differential HI 21-cm brightness temperature ($\delta T_b$) profiles 
around sources at three different stages of the cosmic dawn and reionization. 
In general, the differential brightness temperature of the HI 21-cm signal can be written as
\begin{eqnarray}
 \delta T_b (\vec{\theta}, \nu)  \!\!\!\! & = & \!\!\!\! 27 ~ x_{\rm HI} (\mathbf{x}, z) [1+\delta_{\rm B}(\mathbf{x}, z)] \left(\frac{\Omega_B h^2}{0.023}\right) \nonumber\\
&\times& \!\!\!\!\left(\frac{0.15}{\Omega_m h^2}\frac{1+z}{10}\right)^{1/2}\left[1-\frac{T_{\rm CMB}(z)}{T_{\rm S}(\mathbf{x}, z)}\right]\,\rm{mK},
\nonumber \\
\label{eq:brightnessT}
\end{eqnarray}
where $\mathbf{x} = r_z \mathbf{\hat{n}}$ and $1+z = 1420~{\rm MHz}/\nu_{\rm obs}$, 
$r_z$ is the radial comoving distance to redshift $z$.
$\delta_{\rm B}(z,\mathbf{x})$ and $x_{\rm HI}(z,\mathbf{x})$ denote the density contrast in baryons
and HI fraction  respectively at  
$\mathbf{x}$ at a redshift $z$.  
$T_{\rm CMB}(z)$ = 2.73 $\times (1+z)$ K is the CMB temperature at a redshift $z$
and  $T_{\rm S}$ is the spin temperature of H I gas.
We ignore the line of sight peculiar velocity effects \citep{bharadwaj04, barkana05} 
in the above expression, which, we believe do not affect the results presented here.

Below, we briefly discuss the HI 21-cm brightness temperature profile around individual sources 
expected at different stages of reionization and cosmic dawn.

\subsection{Model A: HI 21-cm signal from ionized bubbles during reionization}

\begin{figure}[h]
\includegraphics[width=12cm]{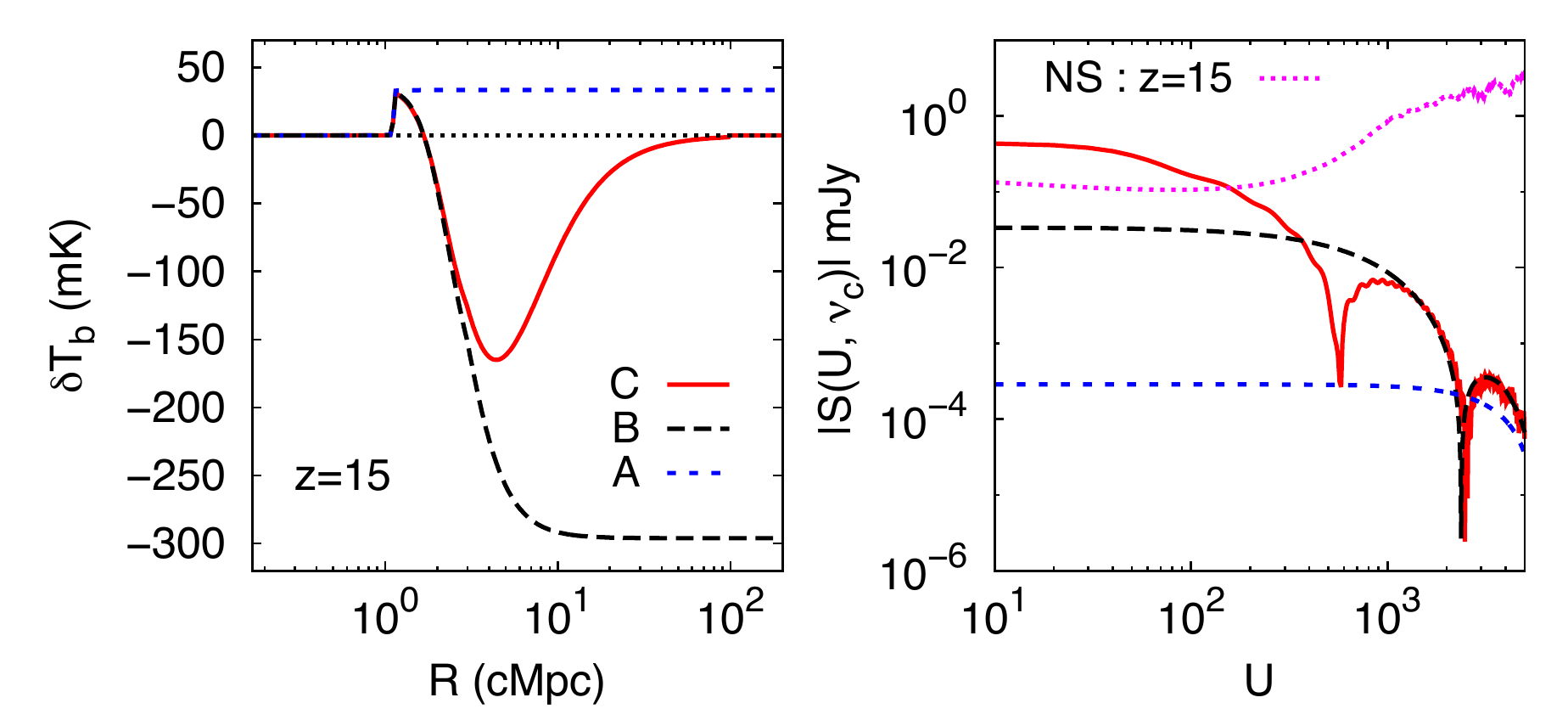}
\centering
\caption{Left panel: The differential brightness temperature profile around an isolated 
mini-QSO. The source properties are taken to be those corresponding to the fiducial values. 
The results are shown for all three coupling models A, B, C described in the texts. 
Right panel: The absolute value of the corresponding visibility amplitude as a function of baseline U.
Also shown are rms noise in the visibilities calculated for $1000$ h of observation 
with the SKA1-low with a frequency resolution of $50 \, {\rm kHz}$. 
The minimum baseline $U_{\rm min}=10$ for the observation we consider here. The figure is taken from
our previous work \cite{ghara15a}.}
\label{fig:signal}
\end{figure}

In this scenario, we assume that the IGM kinetic temperature is fully coupled with the HI spin
temperature and both the temperatures are much higher than the CMB temperature. 
This is a likely scenario when there are sufficient X-ray, Ly-$\alpha$ sources during the cosmic dawn and 
the Universe is already $\sim 10-20\%$ ionized or after that. Now, we consider a spherical 
ionized bubble of comoving 
radius $R_b$ centered at redshift $z_c$ surrounded by uniform IGM with H I fraction
$x_{\rm HI}$. We refer this as model A.  These kind of ionized bubbles are expected to be
common during the later stages
of reionization as there are a significant numbers of UV sources which create spherical 
ionized regions around them.  We note that this model was introduced 
and the formalism for calculating the visibility signal was developed first in \cite{datta07}. 
Here and in the subsection 2.3, we  discuss the key equations for calculating the visibility signal.

A bubble of comoving radius $R_b$ will
be seen as a circular disc in each of the frequency channels that cut
through the bubble. At a frequency channel $\nu$, the angular radius
of the disc is $\theta_{\nu} = (R_b/r_{\nu})\sqrt{1-(\Delta \nu/ \Delta \nu_b)^2}$, where  
$\Delta \nu=\nu-\nu_c$ is the distance from the bubble's centre $\nu_c = 1420 \, {\rm MHz}/(1 + z_c)$
and $\Delta \nu_b = R_b/r'_{\nu}$, $\Delta \nu_b$ is the bubble's radius in frequency space. Here,
$r_{\nu}$ is the comoving distance corresponding to $z = (1420 \, {\rm MHz}/\nu)-1$ and 
$r'_{\nu}= dr_{\nu}/d \nu$.
The specific intensity profile of HI 21-cm signal in this scenario can be written as 
\begin{equation} 
 I_{\nu}(\th)=\bar{I}_{\nu} x_{\rm HI}
\left[1-\Theta\left(1- \frac{\mid \th-\th_c\mid}{\theta_\nu} \right) \right] 
\Theta \left(1- \frac{\mid \nu -\nu_c \mid}{\Delta \nu_b} \right)
\label{eq:I_modelA}
\end{equation}
where the radiation from the uniform background HI distribution with H I fraction 
$x_{\rm HI}$ can be written as 
$x_{\rm HI}$ $\bar{I_{\nu}}$ where $\bar{I_{\nu}}=2.5\times10^2\frac{Jy}{sr} \left (\frac{\Omega_b
  h^2}{0.02}\right )\left( \frac{0.7}{h} \right ) \left
(\frac{H_0}{H(z)} \right )$ and $\Theta(x)$ is the Heaviside step function. Note that, here we
assume $T_S>>T_{\rm CMB}$, H I overdensity parameter $\delta_B=0$. In the Raleigh-Jeans
limit the specific intensity $\bar{I}_{\nu}$ in eq. \ref{eq:I_modelA} can be calculated 
from the differential 
brightness temperature in eq. \ref{eq:brightnessT} using the relation 
$\bar{I}_{\nu}=\left( \frac{2 K_B}{\lambda^2}\right)\delta T_b (\vec{\theta}, \nu)$.

The dashed blue line in the left panel of figure \ref{fig:signal} shows the 
HI differential brightness temperature profile as a function of comoving
distance from the source. As mentioned above, we assume $T_{\rm S}>>T_{\rm CMB}$ here. 
The differential brightness temperature is zero very close to the source because the region 
is highly ionised and equals to the background HI differential brightness temperature far 
from the source.

\subsection{Model B \& C: 21-cm signal around sources during cosmic dawn}
In model B, the Ly-$\alpha$ coupling is assumed to be efficient
throughout i.e, the IGM kinetic temperature is completely coupled with the HI spin temperature. 
Additionally, we assume that there is a X-ray source at the centre of a host DM halo. We refer
this as model B. In this model the HI 21-cm signal pattern, in general,
can be divided into three prominent regions: (i) the signal is
absent inside the central ionized (H II) bubble of the source.
(ii) The H II (ionized) region is followed by a region which is neutral
and heated by X-rays. In this region, the IGM kinetic temperature $T_K>T_{\rm CMB}$ and thus the
HI 21-cm signal is seen in emission. (iii) The third is a strong absorption
region which is colder than $T_{\rm CMB}$ as the X-rays photons have not
been able to penetrate into this region. At reasonably far from the source the spin temperature 
eventually becomes equal to the background IGM temperature.

We consider a third model, referred as model C, which calculates the Ly-$\alpha$ coupling, 
heating, ionisation self-consistently. This is believed to be the case at very early stages
of the cosmic dawn. In addition to all three prominent regions described in the model B 
there will one more prominent region i.e,  Beyond the absorption
region, the signal gradually approaches to zero as
the Ly -$\alpha$ coupling becomes less efficient at the region far away from the
source.

Here, we consider only mini-QSOs as a X-ray sources (see \cite{ghara15a} for other X-rays sources). 
The stellar mass for this source model is taken to be  
$M_{\star} = 10^{7}\, M_{\odot}$.  The value of the escape fraction is taken to be 
$f_{\rm esc}=0.1$. The X-ray to UV luminosity is fixed as $f_X = 0.05$, while the power 
law index of the X-ray 
spectrum of mini-QSO model is chosen to be $\alpha = 1.5$. It is assumed that the age 
of the sources is $t_{\rm age} = 20$ Myr. The densities of hydrogen 
and helium in the IGM are assumed to be uniform and the density contrast $\delta$ is set 
to $0$. We denote this model source as our ``fiducial'' model. We choose the fiducial 
redshift to be $15$ for presenting our results.

The black dashed and the solid red lines in the left panel of the figure \ref{fig:signal} 
show the 21-cm differential brightness temperature  $\delta T_b$ profile around a mini-QSO like 
source described above for the B and C models respectively.

\subsection{Visibilities}

The expected visibility signal in each frequency channel is 
the Fourier transform of the sky intensity pattern and can be approximated  as (see 
\cite{datta07} for the model A, and the Appendix A of \cite{ghara15a} for the B and C models)
\begin{equation}
S^{(i)}(U,\nu) \approx -2m x_{\rm HI} \pi I_{\nu,i}\theta^2_{\nu, i} \left[\frac{J_1(2\pi U \theta_{\nu, i})}{2\pi U \theta_{\nu,i}}\right]\Theta\left(1-\frac{|\nu-\nu_c|}{\Delta\nu_{b,i}}\right).
\label{vis_bubble}
\end{equation} 
We assume that the source is at the phase centre of the antenna field of view. $i$ denotes
various models such as model A, B, C. The integer $m=1$ for 
the A and C model and $-1$ for the B model. The above visibility signal picks up an 
extra phase factor if the source is not at the centre of the field of view. 
The right panel of Fig: \ref{fig:signal} shows the visibility signal $S^{(i)}(U,\nu)$ as a 
function of baseline $U$ for the central frequency channel.

For the model A, the HI 21-cm signal is zero within a circular disc through an 
ionized bubble of radius $R_{\rm b,A}$. 
In each channel of frequency $\nu$, the signal has a peak value 
$|S(0,\nu)|=\pi x_{\rm HI} \bar{I}_{\nu, A} \theta^2_{\nu,A}$.
The signal is largely contained within baselines
$U \le U_0=0.61/\theta_{\nu, A}$ (blue dashed line in the right panel of figure \ref{fig:signal}), 
where the Bessel function has its first zero
crossing, and the signal is much smaller at larger baselines. 

For the B and C model, we find that the emission and absorption
bubbles are larger than the H II bubble. We further find that the visibility signal is essentially
determined by the emission region in the B model and by the absorbtion region in the C model 
(black dashed and red solid lines respectively in the right panel of figure \ref{fig:signal}). 
The first zero crossings of the visibility 
signal for the B and C model occur at lower values of U compared to the model A. For example, the
first zero crossing  appears around $U_0=2390$ and $590$ for the models B and C respectively,
while it appears around $U_0=5650$ for the model A. In addition,
we find that the amplitude of the visibility signal at small U
is the largest (smallest) for C (A). The amplitude of the visibility
at small baselines scales roughly as $\bar{I}_{\nu, i} \theta^2_{\nu,i}$, where $\bar{I}_{\nu, i}$ 
is the signal amplitude in the emission region for the model A
and in the absorption region in models B and C. $\theta_{\nu, i}=R_{\nu,i}/r_{\nu}$ is
the total radial extent 
of the ionized region for the model A, ionized and emission
regions for model B and ionized, emission and
absorption regions for models C. Given this scaling, it is easy to see that since the size of
the absorption region is much larger than the ionized and
emission regions, the visibility amplitude in model C would
be the largest. This is further assisted by the fact that the $\bar{I_{\nu, i}}$ 
itself is very high in the absorption region.

 We would like to mention that the results presented above are for very simplistic models where
ionized bubbles (in case of model A) or heated and absorption (for model B \& C) regions are assumed to be
isolated and, therefore, the signal is not complicated by overlap of other nearby bubbles or 
heated regions. We use numerical simulations where we consider more realistic scenarios such 
as overlap of bubbles, impact of density fluctuations in HI 
distribution  (\cite{datta3, datta12}), effects 
of finite light travel time and varying neutral hydrogen fraction along the line of sight direction
of the bubble (\cite{majumdar12})  during reionization epoch and overlap of heated or absorption
regions during cosmic dawn (\cite{ghara15a}). 

\section{Estimators for detecting sources individually}
The visibility  recorded in radio-interferometric
observations  can be written as a combination of four separate contributions 
\be
V(\u,\nu)=S(\u,\nu)+HF(\u,\nu)+N(\u,\nu)+F(\u,\nu)
\label{eq:vis2}
\e
where the baseline $\u={\bf d}/\lambda$, ${\bf d}$ is physical separation between a pair 
of antennas projected on the plane perpendicular to the line of sight. $S(\vec{U},\nu)$ 
is the HI signal that we are interested in, 
$ HF(\u,\nu)$ is contribution from fluctuating HI outside the targeted  region,
$N(\vec{U},\nu) $ is the system noise which is inherent to the 
measurement and $F(\vec{U},\nu)$ is the contribution from other
astrophysical sources  referred to as the foregrounds. 

It is a major challenge to detect the signal which is expected to be
buried in noise and foregrounds both of which are much stronger. It would be relatively 
simple to detect the signal in a situation where there is only noise and no foregrounds.
There are various ways of reducing the rms noise in observations. Below we discuss two 
ways to minimize the noise contribution in the measurements and maximize the signal 
to noise ratio.

\subsection{Matched filter technique}

 This technique, in the context of ionized bubble detection, 
was first proposed and explored in a simple reionization model in \cite{datta07}. In subsequent works, 
the technique was explored in more realistic reionization scenarios obtained from numerical simulations 
\citep{datta3, majumdar11, datta12, malloy13}). Here we present a brief summary and essential
equations of the matched fiter technique for detecting ionized bubbles.

Bubble detection is carried out by combining the entire observed
visibility signal weighed with the filter. The estimator $\hat{E}$ is defined as \cite{datta07}
\be
\hat{E}=  \left[ \sum_{a,b} S_{f}^{\ast}(\u_a,\nu_b)
\hat{V}(\u_a,\nu_b) \right]/\left[   \sum_{a,b} 1 \right]
\label{eq:estim1}
\e
where the sum is over all frequency channels and baselines. The expectation
value $\langle E \rangle$ is non-zero only if an ionized bubble is present. 
$S_f(\u_a,\nu_b)$ is a filter which has 
been constructed to detect the particular ionized bubble.

The system noise (NS), HI fluctuations (HF) and the foregrounds
(FG) all contribute to the variance of the estimator
\begin{eqnarray}
\langle (\Delta \E)^2 \rangle =\left <(\Delta \hat
E)^2 \right >_{{\rm NS}}+\left<(\Delta \hat E)^2 \right >_{{\rm HF
}} \,
 +\left<(\Delta \hat E)^2 \right >_{{\rm FG}} \,.\nonumber \\ 
\label{eq:16}
\end{eqnarray}

Because of our choice of the matched filter, the contribution from
the residuals after foreground subtraction $\left<(\Delta \hat E)^2 \right >_{{\rm FG}}$
is predicted to be smaller than the signal \citep{datta07, datta12} and we do not consider it in
the subsequent analysis. The contribution $\left<(\Delta \hat E)^2 \right >_{{\rm HF
}}$ which arises from the HI fluctuations outside the target bubble imposes a fundamental
restriction on the bubble detection. It is not possible to detect an
ionized bubble for which $\langle E \rangle \le \sqrt{\left<(\Delta \hat E)^2 \right >_{{\rm HF
}}}$. Bubble detection is meaningful only in situations where the contribution from HI
fluctuations is considerably smaller than the expected signal. In \cite{datta3} we use numerical simulations
to study the impact of HI fluctuations on matched filter search for
ionized bubbles in redshifted 21-cm maps. We find that the fluctuating intergalactic medium 
prohibits detection of small bubbles of radii $ \lesssim 6 $ Mpc and $\lesssim 12$ Mpc for 
the GMRT and the MWA, respectively, however large be the observation time. In a situation
when the contribution from HI fluctuations is negligible the SNR is defined as
\be
SNR=\langle E \rangle /\sqrt{\left<(\Delta \hat E)^2 \right >_{{\rm NS
}}}
\e

It is possible to analytically estimate 
$\langle \E \rangle$ and $\langle (\Delta \hat E)^2 \rangle_{\rm NS}$  in the continuum limit
\cite{datta07}. We have 

\be
\langle \E \rangle  =\int d^2U \, \int d\nu  \, \rho_N(\u,\nu) \, \, 
{S_f}^{\ast}(\u,\nu) S(\u,\nu)  \,,
\label{eq:estim2}
\e 
and 
\be
\langle (\Delta \hat E)^2 \rangle_{\rm NS}
= \sigma^2 
\int d^2U \, \int d\nu  \, \rho_N(\u,\nu) \, \, 
\mid S_{f}(\u,\nu)\mid^2\,.
\label{eq:ns1}
\e 
$\rho_N(\u,\nu)$ is the normalized baseline distribution function
defined so that  $\int d^2U \, \int d\nu \rho_N(\u,\nu)=1$. For
a given observation,  $d^2U \, d\nu\, 
\rho_N(\u,\nu)$ is the fraction of visibilities in the interval  $d^2U \,
d\nu$ of baselines and frequency channels. Further, we expect
$\rho_N(\u,\nu) \propto \nu^{-2}$ for an
uniform distribution of the antenna separations ${\bf d}$.  

The term $\sigma$ in eq. (\ref{eq:ns1}) is the rms.  noise  expected
in an image made using  the radio-interferometric observation being
analyzed. Assuming  observations at two polarizations, we have   
\begin{equation}
\sigma =\frac{k_B T_{sys}}{  A_{eff} \sqrt{N_b t_{obs} B}}
\end{equation}
where $k_B$ is the Boltzmann constant, $T_{sys}$ the system
temperature,  $A_{eff}$ the effective collecting area of an individual
antenna in the array, $N_b$ the number of baselines, $t_{obs}$ the
total observing time and $B$ the observing  bandwidth.  

\subsection{Filter}
In order to detect an ionized bubble whose expected signal is $S(\u,\nu)$
we use the matched filter $S_f(\u,\nu)$ defined as 
\begin{eqnarray}
S_f(\u,\nu) \!\!\!\!\!&=& \!\!\!\!\! \left(\frac{\nu}{\nu_c}\right)^2 \left[ S(\u,\nu) -\right.\nonumber \\
&&\!\!\!\!\! \left. 
\Theta\left(1-2 \frac{\mid \nu - \nu_c \mid}{B'}\right)
\frac{1}{B'} \int_{\nu_c - 
  B'/2}^{\nu_c + B'/2} S(\u,\nu') \, d \nu' \right]. \nonumber \\
\end{eqnarray}
Note that the filter is constructed using the signal that we are 
trying to detect. The term $(\nu/\nu_c)^2$ accounts the frequency dependent
$U$ distribution for a given array.  The function $\Theta$ is the
Heaviside step function. The second term in the square brackets serves to
remove the foregrounds within the frequency range $\nu_c- B'/2$ to
$\nu_c+B'/2$.  Here $B'=4\,\Delta \nu_b$ is the frequency width that
we use to estimate and subtract out a frequency independent
foreground contribution. This, we have seen in Paper I, is adequate to
remove the foregrounds such that the residuals are considerably smaller than
the signal. Further we have assumed that $B'$ is smaller than the total
observational bandwidth $B$. The filter  
$S_f(\u,\nu)$  depends on $[R_b,z_c,\th_c]$ the  comoving  radius, 
redshift and angular position of the target bubble that we are trying
to detect.

\subsection{Combining visibilities}
The success of the matched filter method depends on the ability to find a suitable filter.
The signal to noise ration (SNR) is maximum in a situation when the filter matches perfectly 
with the signal.
Prior knowledge about the signal and its dependence on various parameters 
is necessary in order to find an appropriate filter. The HI 21-cm signal around the 
sources during the cosmic dawn depends on various parameters such as the 
number of Ly-$\alpha$, X-ray, UV photons available, X-ray spectral index, background IGM 
kinetic temperature, source age, UV escape fraction, IGM overdensity etc. 
Dependence of the HI 21-cm signal around cosmic dawn sources on so many unknown 
parameters makes it difficult to choose an appropriate filter for the B and C type 
model. We, therefore, do not attempt to apply the matched filter technique to the B, C
type models. Instead, we simply add the visibility signal from all baselines and frequency
channels to enhance the SNR. This method was first introduced and explored in \cite{ghara15a}. 
The estimator is defined as 
\be
\hat{E_2}=  \left[ \sum_{a,b}  \hat{V}(\u_a,\nu_b) \right]/\left[   \sum_{a,b} 1 \right]
\label{eq:estim3}
\e
Note that the above estimator is a special case of the matched filter estimator i.e, 
$S_{f}(\u,\nu)=1$ for all baselines and frequency channels. The SNR can be written as,
\begin{equation}
{\rm SNR} = \frac{1}{\sigma_N} \frac{\int d^2U \int d \nu ~\rho_N(\u,\nu)~ S(\vec{U}, \nu)}{\int d^2 U \int d \nu ~\rho_N(\u,\nu)},
\label{eq:snr}
\end{equation}
where
\begin{equation}
\sigma_N = \frac{\sqrt{2}~ k_B T_{\rm sys}} {A_{\rm eff}~\sqrt{t_{\rm obs}~B~ N_b}}.
\label{snr_n}
\end{equation}
The quantity $B$ denotes the bandwidth of the observations, and is simply the frequency resolution $\Delta \nu_c$ times the number of frequency channels. 
We refer the readers to \cite{ghara15a} for more details.
 
 \section{Results}
\subsection{Prospects of detecting ionized bubbles (H II regions) during reionization}
\begin{figure}[h]
\includegraphics[width=.6\textwidth,angle=0]{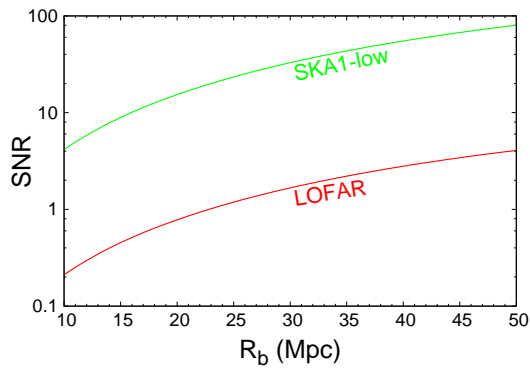}
\centering
\caption{This shows the SNR as a function of comoving radius $R_b$ (Mpc) of ionized bubble using the matched filter 
technique for the model A. Ionized bubbles are assumed to be embedded in uniform IGM with H I
 fraction $x_{\rm HI} \approx 0.5$. The upper and lower lines show results for the SKA1-low and LOFAR
respectively for a total $100$ hr of observations at frequency $165$ MHz corresponding to redshift
$z=7.6$. We consider the SKA1-low baseline distribution given in the paper \cite{ghara15a}. 
Here, we assume that the SKA1-low has a total $512$ number of antennae each with $35$ m diameters.}
\label{fig:SNR}
\end{figure}

In this section we present preliminary results on the detectability of individual ionized bubbles using the 
matched filter technique. We implement this technique only for the model A. Figure \ref{fig:SNR}
shows the SNR as a function of comoving radius $R_b$ (Mpc) of ionized bubble for the matched filter 
technique.  Ionized bubbles are assumed to be embedded in uniform IGM with H I
 fraction $x_{\rm HI} \approx 0.5$. The upper green line of the figure \ref{fig:SNR}
shows results for the SKA1-low  
for a total $100$ hr of observations at frequency $165$ MHz corresponding to redshift
$z=7.6$. For a comparison we show results for LOFAR (red line) for the same observation time.
Here, we assume that the SKA1-low baseline distribution is same as LOFAR. We find  more than $4\sigma$
is possible for the SKA1-low for the entire range of bubble sizes we consider. It is possible to detect
ionized bubbles of radii $R_b=20$ and $30$ Mpc with SNR$>15$ and $30$ with SKA1-low 100 hr 
observations. We also find that the SNR is around $10$ times higher for the SKA1-low compared
to LOFAR. Scaling relations on how the SNR changes with   redshifts, neutral fraction, bubble radius  and  observational parameters can be  found in \cite{datta09}.

\subsection{Prospects of detecting bright QSOs}

\begin{figure}
%\psfrag{xh1}[c][c][1][0]{{\bf {\Large $\xh1$}}}
%\psfrag{tq}[c][c][1][0]{{\bf {\Large $\tau_Q / 10^7\,$ yr}}}

\includegraphics[width=0.6\textwidth,angle=0]{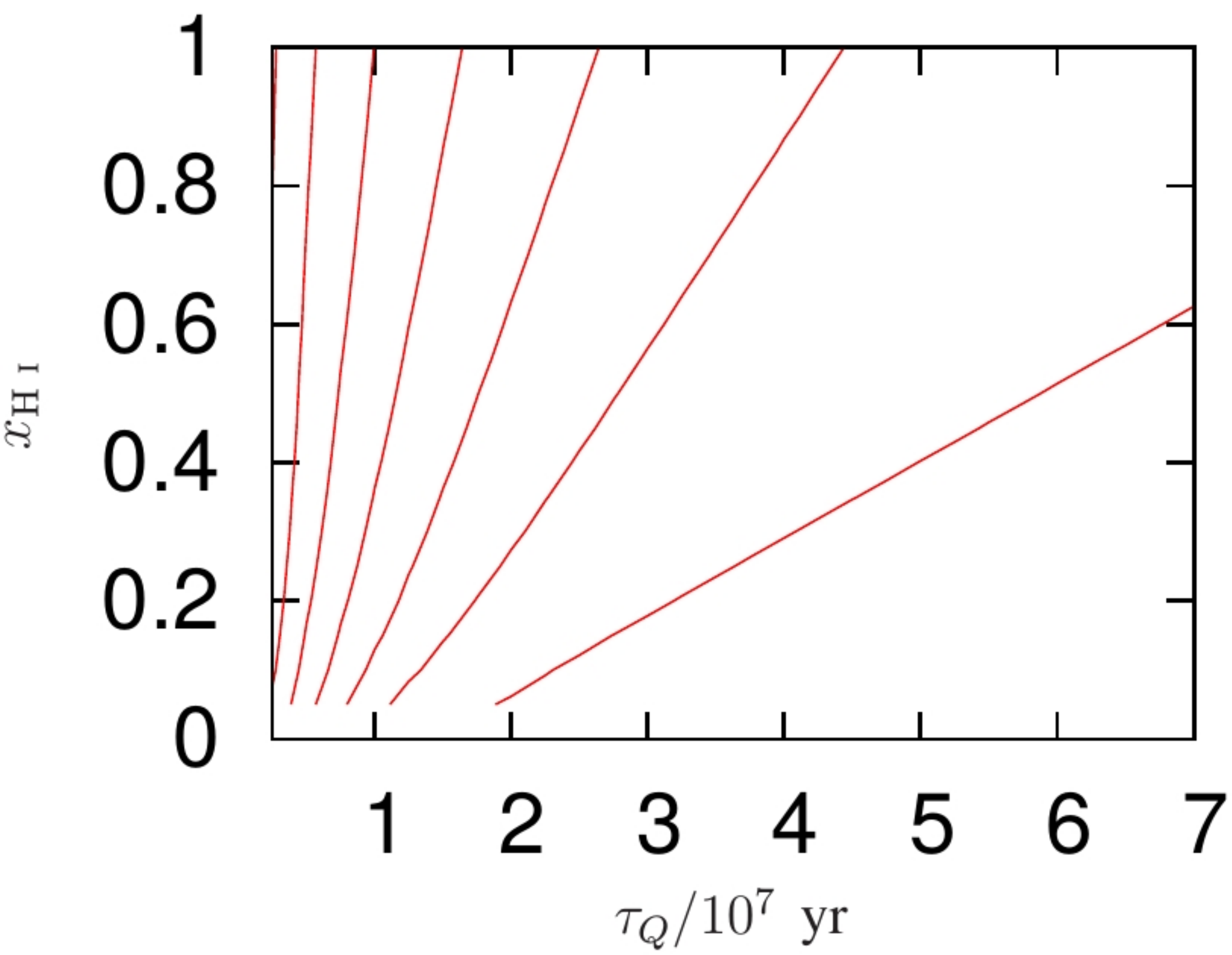}
\centering
\caption{This shows the apparent comoving size of the \HII bubble
  around a quasar with the ionizing photon emission rate
  $\dot{N}_{\gamma} = 1.3 \times 10^{57}\,{\rm sec^{-1}}$, same as
  ULASJ1120+0641 \cite{mortlock11}. Different contours from right to
  left represent $40,\,35,\,30,\,25,\,20,\,15$ and $10$ Mpc.}
\label{fig:bubble_size}
\end{figure}

\begin{figure*}
%\psfrag{xh1}[c][c][1][0]{{\bf {\Large $\xh1$}}}
%\psfrag{tq}[c][c][1][0]{{\bf {\Large $\tau_Q / 10^7\,$ yr}}}
%\psfrag{100hr}[c][c][1][0]{{\textcolor{white}{{\bf {\large $100$
 %         hr}}}}} 
%\psfrag{3sigma}[c][c][1][0]{{{{\bf
 %       {\huge $3\sigma$}}}}} 
%\psfrag{5sigma}[c][c][1][0]{{\bf {\huge
 %     $5\sigma$}}} 
%\includegraphics[width=.4\textwidth,
%  angle=-90]{mort_tobs_5s3s_SKA.ps}
\includegraphics[width=1\textwidth,angle=0]{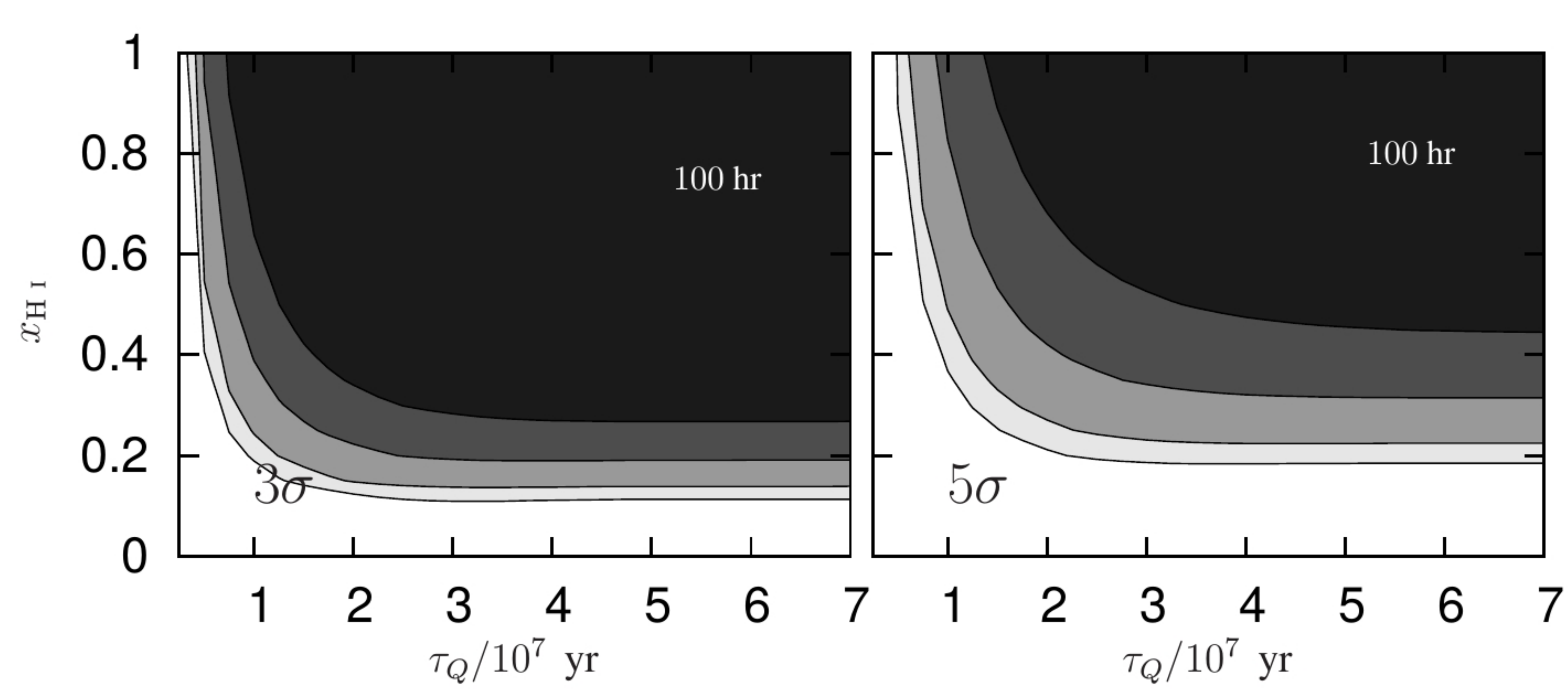}
\centering
  \caption{This shows the estimates (preliminary) of minimum observation
    time required for $3\sigma$ and $5\sigma$ (left and right panels
    respectively) detection of the \HII bubble around the quasar
    ULASJ1120+0641 discovered by \cite{mortlock11} using SKA1-low. 
    Different shades of black from dark to light represent $100$,
    $200$, $400$ and $600$ hr of observations respectively. }
\label{fig:mort_SKA}
\end{figure*}

We estimate the possibility of detecting the \HII bubble around a
quasar using the SKA1-low medium or deep HI 21-cm survey. For this
analysis we assume that the quasar under consideration has a
spectroscopically confirmed redshift from infrared surveys (similar to
the case of \cite{mortlock11}), and its location and the luminosity
(an extrapolation from the measured spectra) are known. We also assume
that the \HII bubble is embedded in uniform IGM with the mean neutral
fraction $\xh1$.  We estimate the total observation time required for
the SKA1-low to detect such an \HII region around a high redshift
quasar using HI 21-cm observations. We investigate the minimum
background $\xh1$ and the quasar age required for a statistically
significant detection of the \HII bubble. We consider the matched
filtering technique described above and introduced in \cite{datta07}
for our analysis. We assume that the quasar under consideration is
emitting with a constant ionizing photon luminosity throughout its age
and, therefore, growing in size following the model of \cite{yu05}
(here we consider it to be same as the \cite{mortlock11} quasar
ULASJ1120+0641 i.e. $\dot{N}_{\gamma} = 1.3 \times 10^{57}\,{\rm
  sec^{-1}}$). We consider the apparent anisotropy in the quasar \HII
regions shape arising due to the finite light travel time \cite{yu05}
and include that as the matched filter parameter in our targeted
search \citep{majumdar11,majumdar12}. The apparent radii of the ionized
bubble for different QSO age and the outside $\xh1$ are shown in
figure \ref{fig:bubble_size}.  We assume that the quasar is located at
redshift $z = 8$.

For the SKA1-low interferometer we
{assume \footnote{https://www.skatelescope.org/key-documents/}}
$A_{eff}/T_{sys} \approx 500\,{\rm m^2\, K^{-1}}$ at $110-160$ MHz
with frequency resolution of $\Delta \nu = 0.1$ MHz, number of
antennas $512$ and a total bandwidth of $32$ MHz, which yields a value
of $C^x = 3.904$ Jy in the following expression of rms of noise
\cite{datta07} contribution in a single baseline
\begin{equation}
\sqrt{\langle \hat{N}^2 \rangle} = C^x \left( \frac{\Delta \nu}{1 {\rm
 MHz}} \right)^{-1/2} \left( \frac{\Delta t}{1 sec}\right)^{-1/2} \,.
\label{eq:noise}
\end{equation}

Using this expression of noise and following the method described in
\cite{majumdar12} we estimate the minimum observation time required
for a $3\sigma$ and $5\sigma$ detection of the \HII bubble using the
SKA1-low. Note that we do not assume the HI  fluctuations outside the
targeted bubble due to galaxy generated \HII regions and density
fluctuations \citep{datta3,datta12,majumdar12}. Figure
\ref{fig:mort_SKA} shows our estimates for SKA1-low. Four different
shades of black from dark to light represent $100$, $200$, $400$ and
$600$ hr of observations respectively.  We find that at least $3
\sigma$ detection is possible with $100$ hr of observations if the QSO
age and the mean $\xh1$ outside the QSO H II region are higher than
$10^7$ Myr and $0.7$ respectively. If the mean $\xh1$ outside the QSO
H II region are $\sim 0.4$, $0.3$ and $0.2$, then the total
observations time required are $200$, $400$ and $600$ hr respectively
when the QSO age is $\sim 10^7$ Myr. A $5 \sigma$ detection is
possible if the QSO age and the outside $\xh1$ are at least $\sim 2
\times 10^7$ Myr and $\sim 0.2$ respectively.

 Note that, results presented above and in subsection 4.1 are for very simplistic models and, therefore, 
should be considered as preliminary one. Detailed investigation needs to be carried out using
realistic simulations of the signal, foregrounds and considering observational issues which we 
aim to address in future.

\subsection{Prospects of detecting sources during cosmic dawn}

\begin{figure*}
\includegraphics[width=12cm]{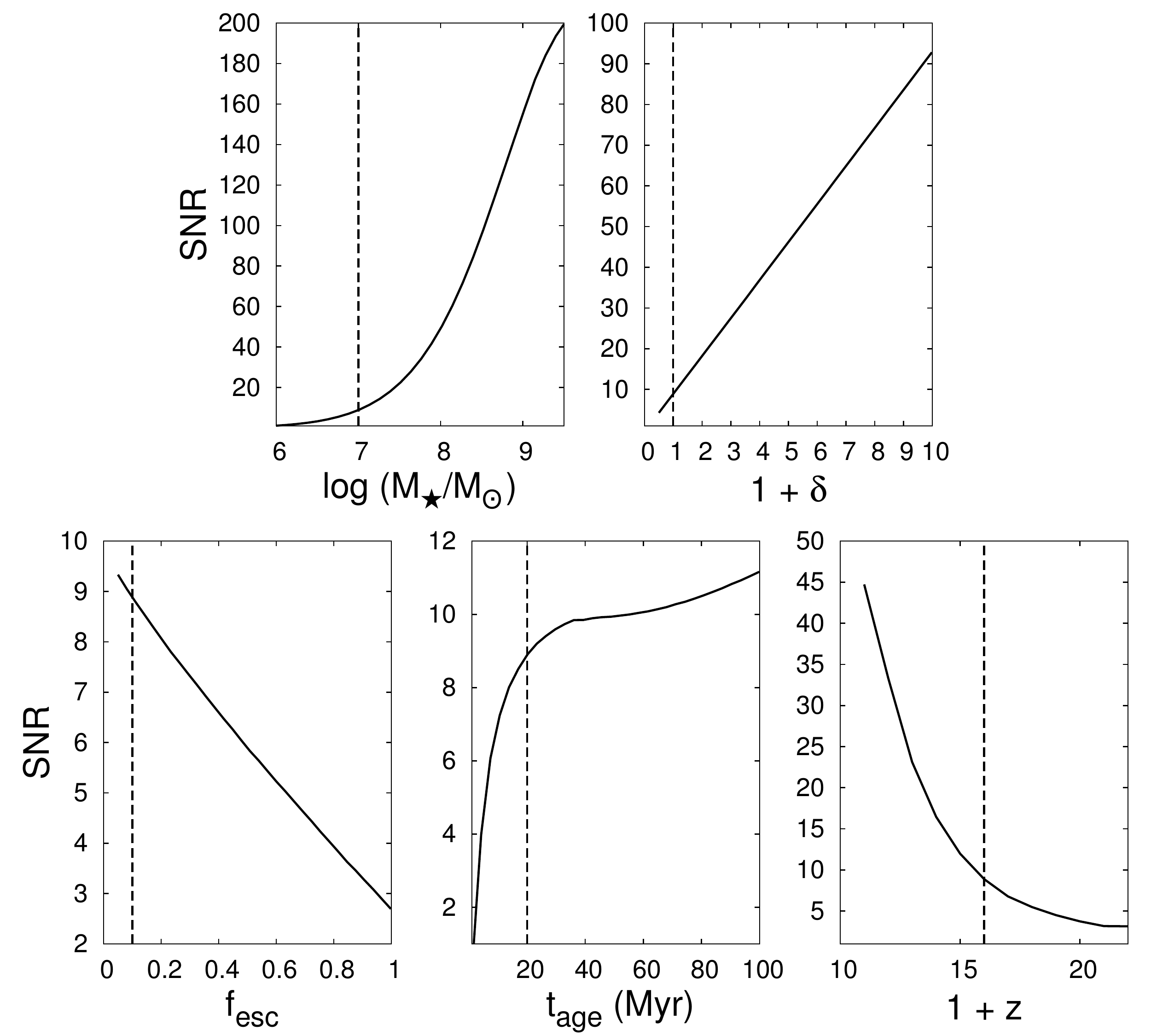}
\centering
\caption{The SNR as a function of different model parameters for the source model 
mini-QSO and coupling model C. While calculating the dependence of the SNR on a particular 
parameter, we have fixed the other parameters to their fiducial values.
The fiducial value for each parameter is denoted by the vertical dashed line in the
 corresponding panel. The SNR is calculated using eq. \ref{eq:snr} for $1000$
hr of observations with SKA1-low with a bandwidth of $16$ MHz. This figure is taken from  \cite{ghara15a}}.
\label{fig:detection3}
\end{figure*}
 In this section, we summarize our results on the prospects of detecting sources during cosmic dawn
( see \cite{ghara15a} for more details).

Different panels in Figure \ref{fig:detection3} show the SNR for the SKA1-low as a function of various parameters 
related to the source properties such as the stellar mass, overdensity 
of the surrounding IGM, escape fraction of ionizing photons, age of the source etc. 
The fiducial values of the parameters are, 
$M_{\star}=10^7 M_{\odot}$, $f_X = 0.05$, $\delta=0$, $f_{\rm esc} = 0.1$,  $\alpha= 1.5$, 
$t_{\rm age} = 20$ Myr. While plotting the dependence of the SNR as a function of a particular parameter, we
have kept all the other parameters fixed to their fiducial values.
The total observational time and bandwidth are $1000$ hours and $16$ MHz respectively.
 We find that that the SKA1-low  should be 
able to detect those sources individually with $\sim 9 \sigma$ significance at redhift $z=15$ 
with  a total observation of $\sim 1000$ hr. 

The strength of the 21-cm signal and corresponding visibility signal increase with the 
increase in the stellar mass, therefore, the SNR increases
with the stellar mass (top left panel).  
We find that (bottom left panel) the SNR decreases with increase of escape fraction.
The amount of Ly-$\alpha$ photons, produced due to 
the recombination in the ISM, is proportional 
to $1 - f_{\rm esc}$. Thus the Ly-$\alpha$ coupling in the IGM becomes weaker 
as $f_{\rm esc}$ increases.
This results in a smaller absorption region and hence a smaller
amplitude of the 21-cm signal. With increase in the age of the source, the size of the 21-cm signal
region increases as the photons propagate longer distance and hence the visibility strength
at lower baselines as well as the SNR increase, as we can see in the bottom middle
panel of Figure \ref{fig:detection3} . As the signal $\delta T_b$
is proportional to the density $(1+\delta)$, the SNR scales linearly with $(1+\delta)$, as
presented in the right top panel. We further note that the SNR decreases
monotonically with increasing redshift which can be understood from the fact
that the system temperature $T_{\rm sys}$ decreases rapidly at lower redshifts. The SNR 
 reduces from $\sim 9$ at $z=15$ to $3$ at $z=20$ for fiducial
values of other parameters. We find that the SNR is weakly dependent on the two X-ray 
parameters i.e, $f_X$ and $\alpha$ and we dont show them here.

\section{Discussion and conclusions} 
Detection of individual luminous sources galaxies, QSOs during the epoch of reionization and cosmic dawn through 
their signature on the HI 21-cm signal is one of the direct approaches to probe the epoch. 
Current low frequency radio telescopes such as the GMRT, MWA, LOFAR primarily aim to detect the HI 21-cm signal
statistically through quantities such as the power spectrum, rms, skewness of the HI 21-cm brightness
temperature fluctuations. However, it has been proposed that these experiments may be
able to detect individual ionized bubbles around reionization sources using optimum detection 
techniques such as the matched filter technique \citep{datta07,datta3,datta09,majumdar11, majumdar12,datta12} 
or in low resolution HI images \citep{geil08,zaroubi12}. With an order of magnitude higher sensitivity, 
better baseline coverage and better instrument the SKA1-low is expected to detect such 
objects individually with much less observations time.

In this paper, we summarize our previous and ongoing works on this issue and present some preliminary
results on the prospects of  detecting individual sources 
during reionization and cosmic dawn using SKA1-low HI 21-cm maps. While calculating the HI
21-cm signal around individual sources we consider three scenarions: (i) The first scenario,
 named model A, assumes that the HI spin
temperature is fully with the IGM kinetic temperature through Ly-$\alpha$ coupling 
 and both the temperatures are much higher than the background CMB temperature. 
This is a likely scenario when the Universe is already $\sim 10-20\%$ ionized or the time period subsequent 
to that. The central source creates a spherical ionized bubble which is embedded in a 
uniform fully neutral (or partially
neutral) medium. (ii) In the second scenario, named model B, we assume that the HI spin temperature is 
completely coupled with the IGM kinetic temperature.  Additionally, we assume that 
the DM halo hosts a X-ray source its centre and calculate the HI spin temperature
and the IGM kinetic temperature profile around it. (iii) Model C which is our third model,
calculates, self-consistently, the Ly-$\alpha$ coupling, heating, ionisation profile around the source. 
This is believed to be the case at very early stages of the cosmic dawn. We discuss various features
 in the differential brightness
 temperature profile around the central source in these three models. Subsequently, we calculate,
 analytically, the corresponding visibility signal and discuss its various features for the above three models.

 We propose two visibility based methods for detecting such signal around individual sources: one
 based on the matched filter formalism we developed \citep{datta07, datta3, datta12} and 
 the other relies on simply combining visibility signal from all 
 baselines and frequency channels (\cite{ghara15a}). We apply the first method to the model A i.e, to spherical
 ionized bubble model during the reionization epoch. We find that the SKA1-low should be 
 able to detect ionized bubbles of radius $R_b \gtrsim 10$ Mpc with $\sim 100$ hr of observations
 at redshift $z \sim 8$ provided that the mean outside neutral fraction $\xh1 \gtrsim 0.5$. 
 Higher observation time would be required for lower neutral fraction. We also investigate 
 the possibility of detecting HII regions around known bright QSOs such as around 
 ULASJ1120+0641 discovered by \cite{mortlock11}.  We find that a $5 \sigma$ detection 
 is possible with $600$ hr of SKA1-low observations 
 if the QSO age and the outside $\xh1$ are at least $\sim 2 \times 10^7$ Myr and 
 $\sim 0.2$ respectively.
 
 Finally, we study the prospects of detecting the very first X-ray and Ly-$\alpha$ sources
  in model B and C during the cosmic dawn. We consider the mini-QSOs as a source of
 X-ray photons. We find that around $\sim 1000$ hr would be required to detect those sources
 individually with SKA1-low. We also study how the SNR changes with various parameters 
related to the source properties such as the stellar mass, escape fraction of ionizing photons, 
X-ray to UV luminosity ratio, age of the source and the overdensity of the surrounding IGM.

Finally, we emphasise that the work we present and discuss here should be treated as a preliminary one.
Further work using detailed simulations needs to be done to understand the signal properties. 
Detail investigation is also necessary in order to understand the impact of the foreground subtraction,
 data calibration, ionospheric turbulence, RFI mitigation effects etc. We plan to adress 
some of these issues in future.

\section*{Acknowledgment}

KKD would like to thank University Grant Commission (UGC), India
for support through UGC-faculty recharge scheme (UGC-FRP) vide ref. no. F.4-5(137-
FRP)/2014(BSR).

\label{lastpage}
\end{document}